# Electronic phase diagram of Cr-doped VO$_2$ epitaxial films studied by *in situ* photoemission spectroscopy

D. Shiga[1,2], X. Cheng[1], T. T. Kim[1], T. Kanda[1,2], N. Hasegawa[1], M. Kitamura[2], K. Yoshimatsu[1], and H. Kumigashira[1,2,*]

[1] *Institute of Multidisciplinary Research for Advanced Materials (IMRAM), Tohoku University, Sendai, 980–8577, Japan*

[2] *Photon Factory, Institute of Materials Structure Science, High Energy Accelerator Research Organization (KEK), Tsukuba, 305–0801, Japan*

**Abstract**

Through *in situ* photoemission spectroscopy (PES), we investigated the changes in the electronic structure of Cr-doped VO$_2$ films coherently grown on TiO$_2$ (001) substrates. The electronic phase diagram of Cr$_x$V$_{1-x}$O$_2$ is drawn by a combination of electric and spectroscopic measurements. The phase diagram is similar to that of bulk Cr$_x$V$_{1-x}$O$_2$, while the temperature of metal-insulator transition ($T_{MIT}$) is significantly suppressed by the epitaxial strain effect. In the range of $x$ = 0–0.04, where $T_{MIT}$ remains unchanged as a function of $x$, the PES spectra show dramatic change across $T_{MIT}$, demonstrating the characteristic spectral changes associated with the Peierls phenomenon. In contrast, for $x$ > 0.04, the $T_{MIT}$ linearly increases, and the metal-insulator transition (MIT) may disappear at $x$ = 0.08–0.12. The PES spectra at $x$ = 0.08 exhibit pseudogap behavior near the Fermi level, whereas the characteristic temperature-induced change remains almost intact, suggesting the existence of local V-V dimerization. The suppression of V-V dimerization with increasing $x$ was confirmed by polarization-dependent x-ray absorption spectroscopy. These spectroscopic investigations reveal that the energy gap and V 3$d$ states are essentially unchanged with $0 \leq x \leq 0.08$ despite the suppression of V-V dimerization. The invariance of the energy gap with respect to $x$ suggests that the MIT in Cr$_x$V$_{1-x}$O$_2$ arises primarily from the strong electron correlations, namely the Peierls-assisted Mott transition. Meanwhile, the pseudogap at $x$ = 0.08 eventually evolves to a full gap (Mott gap) at $x$ = 0.12, which is consistent with the disappearance of the temperature-dependent MIT in the electronic phase diagram. These results demonstrate that a Mott insulating phase without V-V dimerization is stabilized at $x$ > 0.08 as a result of the superiority of Mott instability over the Peierls one.

[*]Author to whom correspondence should be addressed: kumigashira@tohoku.ac.jp

# I. INTRODUCTION

The metal-insulator transition (MIT) of vanadium dioxide ($VO_2$) [1], which is one of the most controversially discussed phenomena for decades, is particularly intriguing because both the structural transition and electron correlation contribute to it [2–20]. The structural phase changes from high-temperature metallic rutile R phase ($P4_2/mnm$) to low-temperature insulating monoclinic $M_1$ phase ($P2_1/c$) across the MIT near room temperature [2,3]. As a result, the first-order MIT is accompanied by an orders-of-magnitude change in conductivity. This phenomenon has become a central topic in modern condensed matter physics for its potential application in prospective electronic devices [21–27]. Further, the unusual phenomenon originating from the interplay of the electron correlation and the lattice provides an opportunity to better understand the underlying physics of strongly correlated oxides.

In the phase transition in $VO_2$, tilting and pairing of V ions along the $c_R$ axis, which is defined as the $c$ axis of the rutile structure in the $M_1$ phase, mark this structural change. Because the distances between paired V ions and between V ion pairs are different, the V ions in $VO_2$ are collectively dimerized along the $c_R$ axis in the $M_1$ phase [2,3]. Although the MIT that is concomitant with the collective dimerization of V atoms is reminiscent of the Peierls transition [4,5], the importance of strong electron correlations in $VO_2$ has also been evident for this MIT from a large number of experimental and theoretical investigations [6,7]. Therefore, the mechanism of the MIT in $VO_2$ is now mainly understood as a cooperative Mott-Peierls (or Peierls-Mott) transition [9–16].

This type of MIT in $VO_2$ has motivated researchers to clarify the role of each instability in the unusual phenomena by changing their balance via physical pressure and/or carrier doping [16–20,23–35]. The doping of acceptor impurities (chemical substitution of $V^{4+}$ by $Al^{3+}$ or $Cr^{3+}$) leads to several insulating phases with similar free energies [17–20,28–32]. In bulk $Cr_xV_{1-x}O_2$, Cr doping produces structural modifications, whereas it exhibits a slight influence on electrical conductivity. For low $x$ ($x < 0.05$), the MIT temperature ($T_{MIT}$) remains almost constant (~340 K), while the insulating monoclinic $M_2$ and triclinic T phases appear instead of the $M_1$ phase. In the $M_2$ phase, only one half of the V atoms dimerizes along the $c_R$ axis, and the other half forms zigzag chains of equally spaced atoms. Meanwhile, in the T phase, the V-V pairs in the $M_2$ phase tilt, and the zigzag chains transform into V-V pairs [17]. In the phase diagram in bulk $Cr_xV_{1-x}O_2$, the T phase occurs between the $M_1$ and $M_2$ phases [17,18,30]. In high $x$ region ($x > 0.05$), $T_{MIT}$ increases almost linearly, and another insulating monoclinic $M_4$ phase becomes dominant, where the collective V-V dimerization appears to vanish [18].

The rich structural phase demonstrates the complicated interplay between the electron correlation and the lattice in the structural and electronic phases of VO$_2$. Therefore, the phase of Cr$_x$V$_{1-x}$O$_2$ is also sensitive to uniaxial or biaxial physical pressure [17,18,20]. In the epitaxial thin films, the lowering of $T_{MIT}$ has been observed as a result of substantial epitaxial strain (biaxial pressure) from the substrates. In the case of VO$_2$ film, the $T_{MIT}$ is lower to ~290 K from ~340 K in bulk when VO$_2$ is coherently grown on TiO$_2$ (001) substrates under in-plane tensile strain that suppresses the dimerization and tilting of V-V pair [35,36]. Furthermore, the $T_{MIT}$ strongly depends on the crystallographic orientation of the substrates, whereas the occurrence of V-V dimerization associated with MIT itself remains unchanged [35,36]. These doping and pressure effects suggest that the complicated electronic phase in Cr$_x$V$_{1-x}$O$_2$ originates from the delicate balance between the instabilities of a bandlike Peierls transition and a Mott transition. Thus, it is interesting to investigate the changes in the electronic phase of Cr$_x$V$_{1-x}$O$_2$ films coherently grown on TiO$_2$ (001) substrates. In particular, an investigation with systematic control of $x$ is crucial to obtain information on the changes of the electronic structures and characteristic dimerization as a function of $x$. However, few studies were conducted on the electronic structure of Cr$_x$V$_{1-x}$O$_2$ in epitaxial film form.

Against this backdrop, in this study, we determined the electronic phase diagram of Cr$_x$V$_{1-x}$O$_2$ films coherently grown on TiO$_2$ (001) substrates. By using the pulsed laser deposition (PLD) method, coherent Cr$_x$V$_{1-x}$O$_2$ films with good crystallinity were obtained. Subsequently, we investigated the changes in the electronic structures and characteristic V-V dimerization of Cr$_x$V$_{1-x}$O$_2$ films via *in situ* photoemission spectroscopy (PES) and x-ray absorption spectroscopy (XAS) measurements. The PES and XAS spectra exhibited remarkable and systematic changes as a function of $x$: (1) The characteristic spectral changes associated with the cooperative Mott-Peierls MIT are almost unchanged up to $x = 0.04$, whereas the hysteresis in the resistivity (indicative of a first-order transition) becomes small with increasing $x$. (2) For $x = 0.08$, the spectral change across the MIT remains intact, whereas a pseudogap seems to be formed at the Fermi level ($E_F$) for the high-temperature (HT) metallic phase. (3) The temperature-driven MIT is accompanied by V-V dimer formation for $0 \leq x \leq 0.08$, whereas the number of V-V dimerization seems to reduce with increasing $x$. (4) The energy gap at the low-temperature (LT) insulating phase is essentially unchanged for $0 \leq x \leq 0.08$ despite the suppression of V-V dimerization. (5) The pseudogap at $x = 0.08$ eventually evolves into a Mott gap for $x = 0.12$, where the temperature-dependent MIT may disappear. The observed invariance of the energy gap with respect to $x$ suggests that the MIT in Cr$_x$V$_{1-x}$O$_2$ arises primarily from the strong electron correlations. These results suggest that the delicate balance between a Mott instability and a bandlike Peierls instability is modulated by carrier doping and epitaxial strain, which consequently induce the complicated electronic phase diagram of Cr$_x$V$_{1-x}$O$_2$ films.

## II. EXPERIMENT

$Cr_xV_{1-x}O_2$ films with thicknesses of approximately 8 nm were coherently grown on the (001) surface of 0.05 wt.% Nb-doped rutile-$TiO_2$ substrates in a PLD chamber connected to an *in situ* PES system at BL-2A MUSASHI of the Photon Factory, KEK [15,16,37,38]. Sintered pellets with appropriate compositions of $(Cr_xV_{1-x})_2O_5$ were used as ablation targets. The films were grown at a rate of 0.02 nm s$^{-1}$, as estimated from the Laue fringes in corresponding x-ray diffraction (XRD) patterns. During the deposition, the substrate temperature was maintained at 400°C and the oxygen pressure was maintained at 10 mTorr. We remark here that we carefully optimized the growth temperature not only to avoid interdiffusion of the constituent transition metals across the interface but also to achieve the coherent growth of $Cr_xV_{1-x}O_2$ films with high crystallinity [15]. The surface structures and cleanness of the films were confirmed via reflection high-energy electron diffraction and core-level photoemission measurements, respectively. The detailed characterization results for the grown $Cr_xV_{1-x}O_2$ films are presented in the Supplemental Material [39].

The surface morphologies of the measured films were analyzed via atomic force microscopy in air (see Fig. S1 in the Supplemental Material [39]). The crystal structure was characterized by XRD, which confirmed the coherent growth of single-phase $Cr_xV_{1-x}O_2$ films over the entire composition range of $0 \leq x \leq 0.12$. Furthermore, a sharp diffraction pattern with well-defined Laue fringes was clearly observed, indicating the achievement of solid solution in the films and the formation of homogeneously coherent films with atomically flat surfaces and chemically abrupt interfaces (see Fig. S2 in the Supplemental Material [39]). The electrical resistivity was measured with a temperature ramp rate of 10 mK s$^{-1}$ using the standard four-probe method.

PES measurements were performed *in situ* with the use of a VG-Scienta SES-2002 analyzer with total energy resolutions of 120 meV and 200 meV at photon energies of 700 eV and 1200 eV, respectively. The vacuum-transferring of the grown samples was necessary to prevent the overoxidation of the surface layer [15]. The XAS spectra were also measured *in situ* with linearly polarized light via the measurement of the sample drain current. For linear dichroism (LD) measurements of oxygen *K*-edge XAS (O *K* XAS), we acquired the XAS spectra at angles of $\theta = 0°$ and $60°$ between the $a_R$-axis direction ($[100]_R$) and the polarization vector $\boldsymbol{E}$ while maintaining a fixed angle between the direction normal to the surface and the incident light (see Fig. S3 in the Supplemental Material [39]). The $E_F$ of each sample was determined by the measurement of a gold film that was electrically connected to the sample. As it is common knowledge that $VO_2$ exhibits MIT upon irradiation by light [43], we paid particular attention to the possible spectral change induced by light irradiation. The stoichiometry of the samples was carefully characterized by analyzing the relative intensities of the relevant core levels, confirming that the cation composition of the samples was the same as that of the ablation targets.   Furthermore, the Cr ions were doped as trivalent ions and formed localized Cr 3$d$ states at binding energies of 1.5–1.8 eV, as confirmed by Cr $L$-edge XAS (Cr $L$ XAS) and Cr 2$p$–3$d$ resonant PES measurements (see Fig. S4 in the Supplemental Material [39]).

# III. RESULTS

## A. Electronic phase diagram of $Cr_xV_{1-x}O_2$ films

Figure 1 shows the temperature-dependent resistivity ($\rho$–$T$) curves upon cooling and heating for $Cr_xV_{1-x}O_2$ films grown on nondoped $TiO_2$ substrates under identical growth conditions. For a $VO_2$ film, the typical behaviors across the MIT are observed; the $\rho$–$T$ curve steeply changes across the MIT accompanied by the thermal hysteresis characteristic of a first-order phase transition [15,35,44]. The $T_{MIT}$ is determined to be 289 K, which is defined as the average value of $T_{MIT}$s during cooling ($T_{cool}$ = 286 K) and heating ($T_{heat}$ = 291 K) [44]. The change in resistivity across the MIT [$\rho$(250 K)/$\rho$(320 K)] is considerably larger than $10^3$. These values are almost the same as the corresponding values of previously reported epitaxial $VO_2$ films grown on $TiO_2$ (001) substrates under in-plane tensile strain [15,35,44], which guarantees that the qualities of $VO_2$ films in this study are comparable to those in the previous studies. Furthermore, the abrupt change in the $\rho$–$T$ curve across the MIT suggests that the influence of the interdiffusion of the constituent transition metals across the interface is negligible in the present sample [15].

As $x$ increases, $T_{MIT}$ itself remains almost unchanged up to $x$ = 0.04, although the change in resistivity and the width of the hysteresis across the MIT are significantly reduced. The characteristic $\rho$–$T$ behavior of $x$ = 0 is preserved up to $x$ = 0.02, while it significantly weakens at $x$ = 0.04, where the $\rho$–$T$ curve exhibits a relatively broad MIT with a concomitant small but detectable hysteresis loop. In contrast to the invariance of $T_{MIT}$ in the range of $x$ = 0–0.04, it increases linearly to 313 K at $x$ = 0.06 and 328 K at $x$ = 0.08 beyond the original $T_{MIT}$. For $x$ = 0.06, the small but detectable hysteresis loop exists as in the case of $x$ = 0.04. Meanwhile, for $x$ = 0.08, the $\rho$–$T$ curve exhibits a broad MIT without any characteristic hysteresis behavior, implying that the collective (bandlike) Peierls transition no longer occurs. Eventually, the MIT itself seems to disappear for $x$ > 0.12 because the $\rho$–$T$ curves do not exhibit any kink structures, although the possibility for the occurrence of the MIT above the measurement temperature range is not completely dismissed.

The results are summarized as the phase diagram in Fig. 1(b). The obtained electronic phase diagram of $Cr_xV_{1-x}O_2$ films mimics that of the bulk [17,18]. Although the $T_{MIT}$s reduce by approximately 50 K owing to the strain effect that suppresses the dimerization and tilting of the V-V pair, the characteristic features of the bulk electronic phase are maintained: the $T_{MIT}$ remains constant up to a certain value of $x$ and then increases linearly, accompanied by the reduction of the change in resistivity and the width of the hysteresis across the MIT [18]. The behavior is in sharp contrast to that observed in the previous studies on $Cr_xV_{1-x}O_2$ films that are not coherently grown onto substrates [28–30,32]; in previous fully-relaxed $Cr_xV_{1-x}O_2$ films, $T_{MIT}$ was almost the same as the bulk one for $VO_2$ ($x$ = 0) and monotonically increased with increasing $x$ [32].

The similarities of the electronic phase diagram between the present coherent films and the bulk of $Cr_xV_{1-x}O_2$ [17,18] suggest that the high-quality epitaxial films synthesized in this study exhibit intrinsic behavior. A more detailed inspection of the electronic phase diagram reveals the difference in hysteresis behavior across MIT between the film and bulk, which is responsible for the V-V dimerization. In the bulk, the hysteresis across the MIT still existed even at $x = 0.12$, whereas $T_{MIT}$ significantly increased beyond the original value [18]. In contrast, in the epitaxial films, the hysteresis almost vanishes at $x = 0.06$–$0.08$, suggesting the importance of strain effects and resultant V-V dimerization modulation. The absence of hysteresis for $x > 0.06$ may reflect the fact that the collective Peierls transition no longer plays a vital role in insulating states of $Cr_xV_{1-x}O_2$ films with $x > 0.06$.

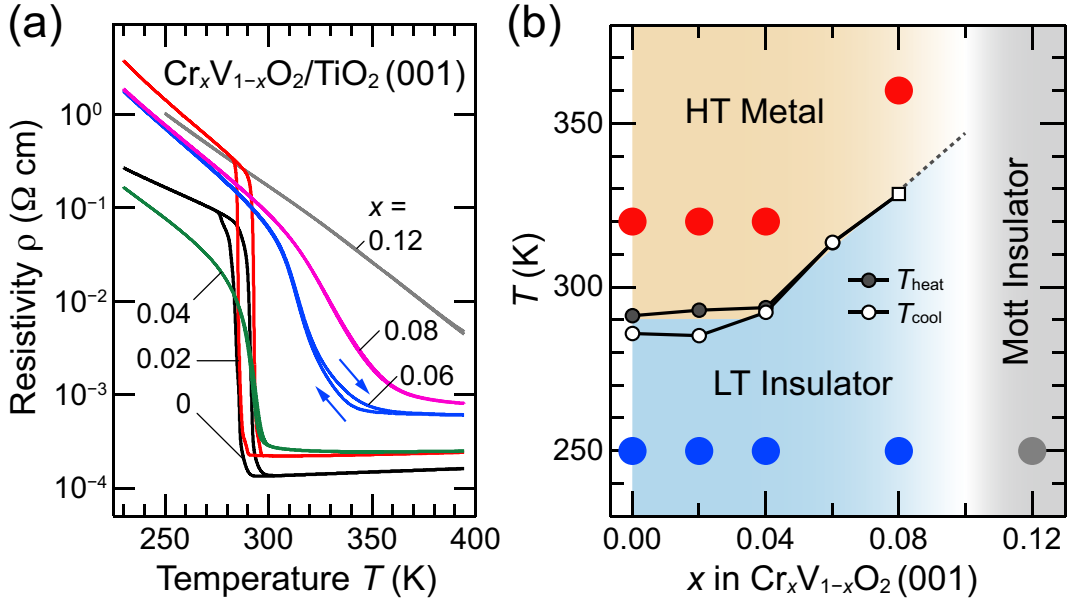

FIG. 1. (a) Temperature dependence of resistivity ($\rho$–$T$) for epitaxial $Cr_xV_{1-x}O_2$/$TiO_2$ (001) films with $x$ ranging from 0 to 0.12. (b) Electronic phase diagram of $Cr_xV_{1-x}O_2$ (001) films as functions of temperature and $x$. The small solid and open circles indicate $T_{heat}$ and $T_{cool}$, respectively, as determined from the $\rho$–$T$ curves. Here, $T_{heat}$ and $T_{cool}$ are defined as the inflection points in $\log_{10}\rho$–$T$ curves upon heating and cooling, respectively. Since the $\rho$–$T$ curve for $x = 0.08$ exhibits no detectable hysteresis loop, its $T_{MIT}$ is indicated by an open square. The solid and dashed lines are merely guides for ease of visualization. Note that MIT no longer occurs for $x = 0.12$. The HT metallic, LT insulating, and Mott insulating phases are assigned from the spectroscopic results as discussed in the manuscript. Large colored solid circles indicate spectroscopic measurement points.

## B. Electronic structure determined by photoemission spectroscopy

Figure 2 shows the temperature dependence of the valence-band spectra of $Cr_xV_{1-x}O_2$ films grown on Nb:TiO$_2$ (001) substrates with varying $x$. The spectra are taken at the points shown in Fig. 1(b). These spectra mainly contain two features: structures derived from O 2$p$ states at binding energies of 3–10 eV and peaks derived from the V 3$d$ states near $E_F$ [11,15,45–47]. Note that the Cr 3$d$ states are localized at 1.5–1.8 eV as trivalent states, which was confirmed by Cr 2$p$–3$d$ resonant PES and Cr $L$ XAS measurements [39]. Regarding the VO$_2$ films, the spectra exhibit the characteristic features representative of the MIT of VO$_2$ [11,45–47]; the spectrum near $E_F$ in the HT metallic phase ($T$ = 320 K) consists of a sharp coherent peak just at $E_F$ and a weak broad satellite structure around 1.2 eV, while that in the LT insulating phase ($T$ = 250 K) exhibits a single peak around 0.8 eV, which leads to the formation of an energy gap at $E_F$. Furthermore, focusing on the O 2$p$ states, we observe dramatic changes across the MIT. These spectral changes are responsible for the structural change concomitant with the MIT in VO$_2$ [11,15].

The spectral changes across the temperature-dependent MIT are clearly observed in the range of $x$ = 0–0.08, as can be seen in Fig. 2(a). Intriguingly, even when $x$ increases up to 0.08, the characteristic temperature-dependent features near $E_F$ representative of the MIT in VO$_2$ remain almost intact. To investigate the spectral changes near $E_F$ in more detail, we present the spectra near $E_F$ according to the enlarged binding-energy scale in Fig. 2(b) for the HT metallic phase ($T > T_{MIT}$) and the LT insulating phase ($T < T_{MIT}$). For the LT insulating phase, the spectrum of $x$ = 0.12 is overlaid for comparison, which is discussed later. As can be seen in Fig. 2(b), the line shapes of V 3$d$ states near $E_F$ are almost identical over the range of $x$ = 0–0.08, whereas there is a slight reduction in the intensity just at $E_F$ for the HT metallic phase of $x$ = 0.08. Especially, the energy gap size and spectral shape of V 3$d$ states in the LT insulating phase are almost the same irrespective of dramatic change in the $\rho$–$T$ curves in this composition range (see Fig. 1), suggesting a common origin for the LT insulating phase.

In contrast to the identical behavior in the LT insulating phase, the PES spectra for the HT metallic phase at $x$ = 0.08 show the intensity reduction in the V 3$d$ derived coherent peak at $E_F$. In addition, the leading edge of the V 3$d$ states shifts from above $E_F$ to below $E_F$ at $x$ = 0.08, suggesting the evolution of a pseudogap at $E_F$. Meanwhile, the characteristic temperature-induced spectral change in V 3$d$ states near $E_F$ is still visible even at $x$ = 0.08, irrespective of the pseudogap formation. The pseudogap may be responsible for the disappearance of the hysteresis in the corresponding $\rho$–$T$ curves, implying the existence of local V-V dimerization [15,48] in this composition region.

With further increasing $x$ to 0.12, the sharp V 3$d$ states near $E_F$ collapse to broad states centered at 1 eV and form an energy gap at $E_F$. The existence of broad V 3$d$ states reminiscent of the lower

Hubbard band in $x = 0.12$ is consistent with previous results [15,38].　　As shown in the upper panel of Fig. 2(b), the spectrum of $x = 0.12$ is markedly different from the others, indicating that $x = 0.12$ is in a different electronic phase.　　Therefore, it is considered that the pseudogap at $x = 0.08$ evolves to a full gap (Mott gap) at $x = 0.12$ through the composition-derived MIT.　　The absence of the spectral weight at $E_F$ for $x = 0.12$ may be responsible for the absence of any kink structures, i.e., MIT behavior, in the corresponding $\rho$–$T$ curve.　　These results suggest that the $Cr_xV_{1-x}O_2$ films with $x > 0.08$ transform into a Mott insulator without the V-V dimerization.

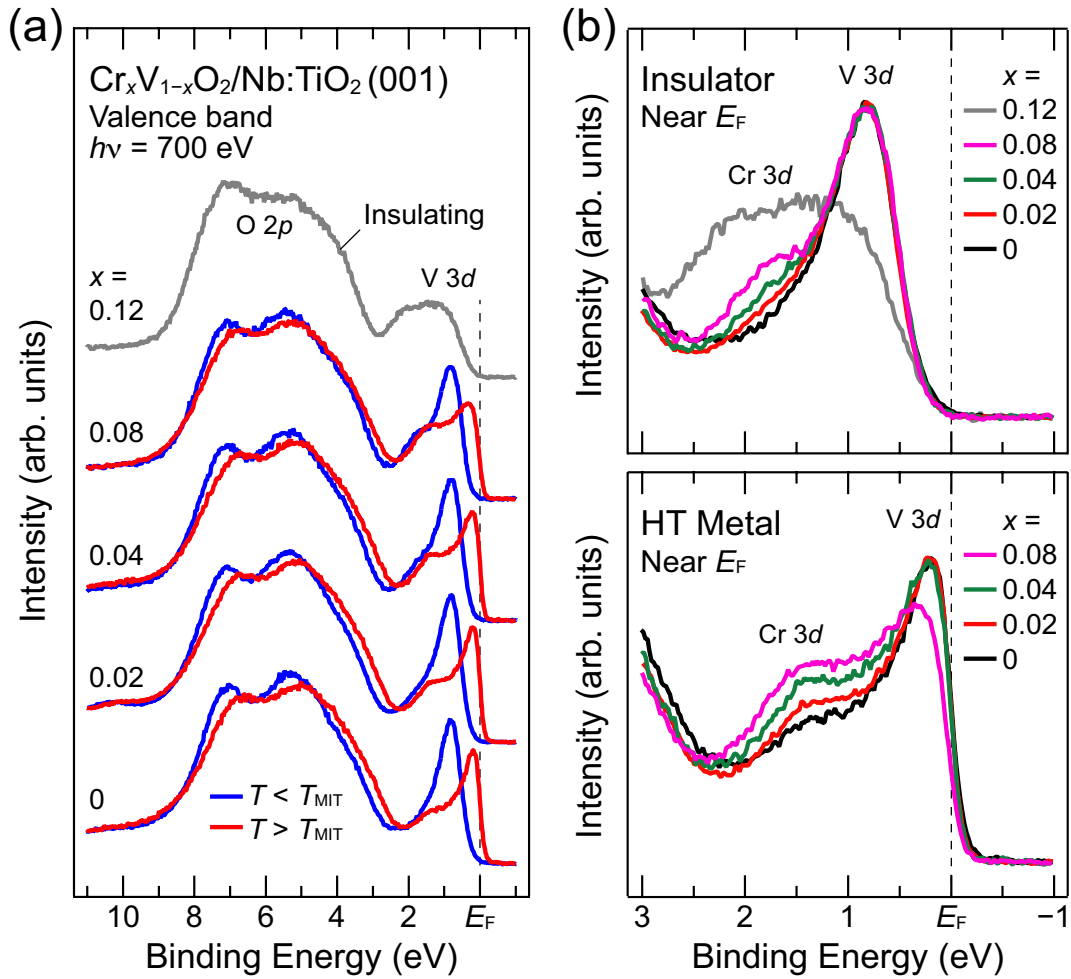

FIG. 2. (a) Valence-band spectra measured at $T < T_{MIT}$ (LT insulating phase) and $T > T_{MIT}$ (HT metallic phase) for the epitaxial $Cr_xV_{1-x}O_2$/Nb:$TiO_2$ (001) films with $x$ ranging from 0 to 0.08, in addition to that at $T = 250$ K for $x = 0.12$. (b) Near-$E_F$ spectra in an expanded energy scale for the insulating phases (upper panel) and the HT metallic phase (bottom panel). Note that the colors of each spectrum correspond to those in Fig 1(a).

**C. V-V dimerization studied by oxygen *K*-edge x-ray absorption spectroscopy**

With respect to the O 2*p* states in the energy range of 3–10 eV [Fig. 2(a)], dramatic changes across the MIT are observed, but they gradually weaken with increasing *x*. Even in the case of *x* = 0.08, weak but distinct changes are observed across the MIT, irrespective of the absence of hysteresis in the $\rho$–*T* curve. Meanwhile, for *x* = 0.12, the O 2*p* states seem to be different from those of the other insulating phases. The crossover of some kind from *x* = 0–0.08 to 0.12 may be related to the significant change in V 3*d* states near $E_F$ [Fig. 2(b)], indicating that another insulating phase emerges at *x* > 0.08 [15,38]. Because the temperature-induced changes in the O 2*p* states are responsible for the structural changes accompanied by the MIT in VO$_2$ [11,15], these results strongly suggest that such a structural change weakens with increasing *x* and may disappear at *x* = 0.08–0.12. To elucidate the crossover of the structural transition (characteristic V-V dimerization of VO$_2$), we have measured the polarization dependence of O *K* XAS, which has previously been utilized as an adequate indicator of V-V dimerization [11,38,49–51], as shown in Fig. 3. The O *K* XAS is a technique complementary to PES for investigating the electronic structures in conduction bands via the probing of the unoccupied O 2*p* partial density of states that are mixed with the unoccupied V 3*d* states. Because the V-V dimerization splits a half-filled $d_{//}$ state into occupied $d_{//}$ and unoccupied $d_{//}^*$ states [3], an additional peak corresponding to the $d_{//}^*$ state appears in the XAS spectra. Furthermore, owing to the strict dipole selection rule, the additional $d_{//}^*$ states only appear in the spectra acquired with the polarization vector ***E*** parallel to the $c_R$ axis (***E*** // $c_R$). As can be confirmed from the XAS spectra of VO$_2$ shown in Fig. 3, the $d_{//}^*$ peak emerges at 530.6 eV (as indicated by the solid triangle) in the M$_1$ (LT insulating) phase (*T* = 250 K) measured at the ***E*** // $c_R$ geometry, whereas it disappears in the R (HT metallic) phase (*T* = 320 K) [Fig. 3(a)]. Furthermore, the identification of the $d_{//}^*$ states is confirmed by inferring the polarization dependence (i.e., LD) of the XAS spectra; the additional $d_{//}^*$ peak in the M$_1$ (LT insulating) phase disappears for the spectrum taken with ***E*** perpendicular to the $c_R$ axis (***E*** ⊥ $c_R$) [Fig. 3(b)]. Thus, the existence of the $d_{//}^*$ peak in the spectra at the ***E*** // $c_R$ geometry can be used as a fingerprint of the V-V dimerization in Cr$_x$V$_{1-x}$O$_2$.

As can be seen in Fig. 3, the $d_{//}^*$ states appearing at the M$_1$ (LT insulating) phase for *x* = 0 begin to weaken with increasing *x* and may disappear at *x* = 0.12. The composition dependence is consistent with that of the O 2*p* states in valence-band structures [Fig. 2(a)]. The weakening of $d_{//}^*$ states while maintaining other states ($\pi^*$ states at 529.4 eV and $\sigma^*$ 531.9 eV) suggests that the number of V-V pair merely decreases with increasing *x*. Moreover, these spectroscopic results imply that the V-V dimerization no longer occurs for *x* = 0.12; in other words, it would be in another insulating phase. The compositional-driven transition in the insulating phase is further supported by the leading-edge shift in the XAS spectra [Fig. 3(a)], which is the counterpart of the energy-gap formation in the PES results in Fig. 2(b); the energy position of the edge for *x* = 0.12 shifts slightly to a higher photon-energy side than other values of $x$ by 0.2 eV. The emergence of the different insulating phase associated with the disappearance of the V-V dimerization is responsible for the broad localized V $3d$ states in the PES spectrum (Fig. 2). Thus, in connection with the PES results, it is naturally concluded that the LT insulating phase for $x = 0$–0.08 changes into another Mott insulating phase without the characteristic V-V dimerization.

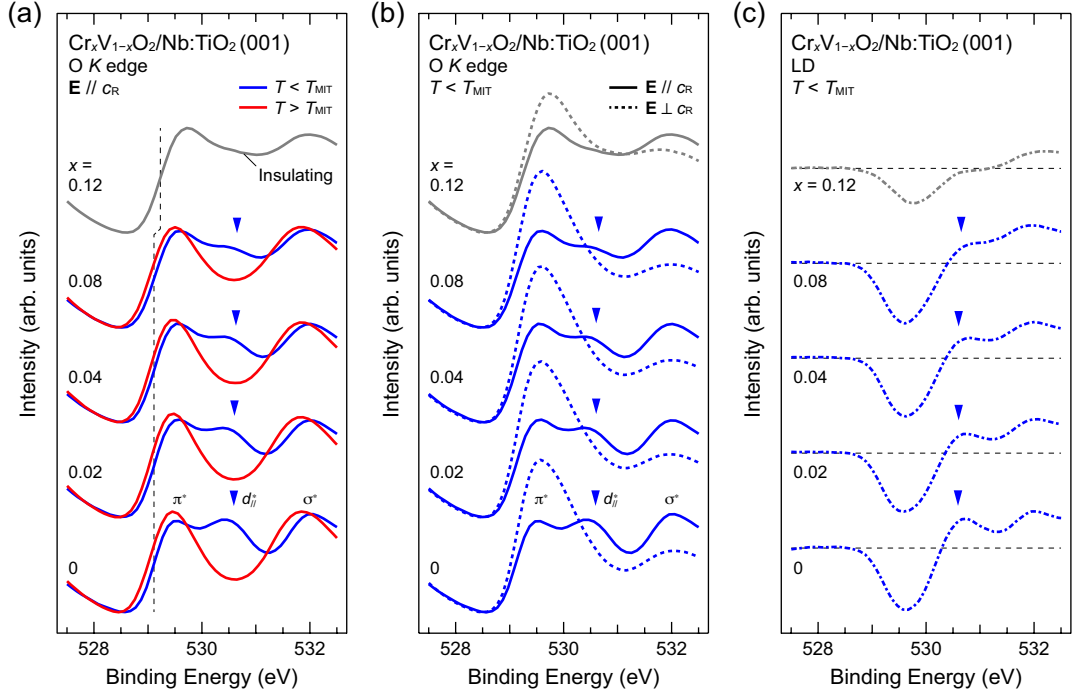

FIG. 3. (a) O $K$ XAS spectra acquired with $E \parallel c_R$ geometry (see Fig. S3 in the Supplemental Material [39]) at $T < T_{MIT}$ (LT insulating phases; bule) and $T > T_{MIT}$ (HT metallic phases; red) of the epitaxial $Cr_xV_{1-x}O_2/Nb:TiO_2$ (001) films with $x$ ranging from 0 to 0.08, in combination with that for $x = 0.12$ at $T = 250$ K (the other insulating phase; gray). Following the assignments made in previous studies [49], the first peak around 529.4 eV and second peak around 531.9 eV can be assigned to $\pi^*$ bands formed by V $3d_{xz}$ and $3d_{yz}$ orbitals and $\sigma^*$ bands formed by $3d_{3z^2-r^2}$ and $3d_{x^2-y^2}$ orbitals, respectively. The additional peak that emerges around 530.6 eV for the spectra acquired at $T < T_{MIT}$ (M$_1$ phase for $x = 0$) can be assigned to the $d_{\parallel}^*$ state due to V-V dimerization (indicated by solid triangles). (b) XAS spectra for insulating phases acquired with $E \parallel c_R$ (solid lines) and $E \perp c_R$ (dashed lines). The XAS spectra with $E \parallel c_R$ ($I_{\parallel}$) are deduced from the expression $I_{\parallel} = (4/3)(I - I_{\perp}/4)$, where $I$ and $I_{\perp}$ (namely, that corresponding to $E \perp c_R$) denote XAS spectra measured with grazing ($\theta = 60°$) and normal ($\theta = 0°$) incidences, respectively. (c) Corresponding LD spectra obtained from the polarization-dependent XAS spectra shown in (b). Solid triangles indicate the position of shoulder structures originating from the $d_{\parallel}^*$ states due to V-V dimerization.

# IV. DISCUSSION

Here, we discuss the origin of the complicated electronic phase (composition dependence of $T_{MIT}$) in $Cr_xV_{1-x}O_2$ films in terms of the roles of Mott and Peierls instabilities. The $x$-dependent physical properties of $Cr_xV_{1-x}O_2$ should be responsible for the delicate balance between the Mott and Peierls instabilities [9–16]. As $x$ increases, there are two effects that may diminish the cooperative Mott-Peierls (or Peierls-Mott) transition: the reduction of a bandlike Peierls instability due to hole doping (chemical substitution of $V^{4+}$ with $Cr^{3+}$) [17–19,28–30] and disorder effects concomitant with the chemical substitution [52,53]. The latter causes a reduction in the effective number of neighbor V ions in a one-dimensional V chain along the $c_R$ axis, in addition to a stabilization of insulating states through the strong localization. According to the Peierls theorem, the two effects considerably suppress the bandlike Peierls instability, and slight hole doping may cause an abrupt decrease in $T_{MIT}$. In particular, such behavior is ubiquitous for conventional low-dimensional materials [52,53]. In contrast to the prediction from the Peierls theorem, the $T_{MIT}$ in $Cr_xV_{1-x}O_2$ films remains unchanged for $x = 0$–$0.04$ and rather increases with increasing $x$ up to $0.08$. Meanwhile, the Mott instability may remain almost unchanged because the on-site Coulomb repulsion $U$ is not sensitive to changes in the environment of V ions owing to the narrow spatial distribution of V $3d$ electrons. From the spectroscopic results, the energy gap is essentially unchanged in the range of $x = 0$–$0.08$, despite the suppression of V-V dimer formation. As the energy gap is insensitive to the different V-V dimerization, V-V dimerization is not considered as the dominant contribution to the energy gap formation. The invariance of the energy gap with respect to $x$ suggests that the energy-gap size arises primarily from the strong electron-electron correlations.

For the spectroscopic results on the HT metallic phase, although the overall electronic structures remain unchanged in the range of $x = 0$–$0.08$, the pseudogap (suppression of spectral weight at $E_F$) evolves at $x = 0.08$. The pseudogap may originate from the disorder effect due to the chemical substitutions. Indeed, similar pseudogap formation has been observed in the coherent states of disordered strongly correlated oxides [54–58]. Nevertheless, the characteristic spectral changes occur across the MIT even at $x = 0.08$. These results suggest that the local V-V dimer formation still plays a critical role in the MIT. Disordered local dimer formation may cause the local dimer-assisted Mott transition; however, it is not a collective transition and consequently is not observed in the $\rho$–$T$ measurements as thermal hysteresis. Therefore, the MIT with a transition over a broad temperature is observed in the $\rho$–$T$ curve for $x = 0.08$, whereas the PES spectra exhibit a change similar to a cooperative Mott-Peierls transition in $VO_2$.

The pseudogap observed at $x = 0.08$ evolves into an energy gap at $x = 0.12$. The spectra of $x = 0.12$ significantly differ from those of other $x$ and are reminiscent of disordered Mott insulators [54–58].

This indicates that the composition-derived MIT in $Cr_xV_{1-x}O_2$ films is predominantly governed by the strong disorder due to chemical substitution. In $Cr_xV_{1-x}O_2$ with $Cr^{3+}$ and $V^{4+}$ ions randomly occupying the transition metal sites, the $Cr^{3+}$ ions perturb the periodic potential of the V $3d$ band, which introduces significant disorder and resultant Anderson-localized states [59–62]. Meanwhile, $Cr_xV_{1-x}O_2$ is a strongly correlated electron system. Thus, the disorder-induced localization is enhanced by the interplay with strong electron-electron interactions. This is consistent with the absence of V-V dimerization and resultant composition-derived MIT at $x = 0.12$. The observed spectral behavior and composition-dependent MIT can be understood in terms of the combined effects of electron correlations and disorder potentials.

Finally, we briefly discuss the epitaxial strain effect on $Cr_xV_{1-x}O_2$ films. Compared with the electronic phase diagram of the bulk [17,18], that of $Cr_xV_{1-x}O_2$ in epitaxial film form has similar features, although the $T_{MIT}$ is lower by 50 K. Previous studies on bulk $Cr_xV_{1-x}O_2$ suggest the formation of complicated dimer structures as functions of $x$ and temperature [17,18]. As in the Cr-doped $VO_2$ system at the transition from R to $M_2$, T, or $M_4$ phases, the V lattice splits into two sublattices [18,53]. The evolution of electronic structures with $x$ has been interpreted as a two-band model, which is phenomenological considering the two electronic states corresponding to the two sublattices [18,53]. However, the present results on the $Cr_xV_{1-x}O_2$ films differ from the electronic structure predicted from the model. As shown in Fig. 3, the $d_{//}^*$ states are always located between $\pi^*$ and $\sigma^*$ states and merely reduce their intensities with increasing $x$. The counterpart of the $d_{//}$ states, which are clearly observed in the PES spectra at the LT insulating phase (Fig. 2), do not change its energy position in the range of $x = 0$–$0.08$. These results suggest that the V-V dimer structure locally collapses with increasing $x$, and the phases with characteristic dimer structures in bulk are absent in the films, with the exception of the $M_1$ phase in the $VO_2$ film. According to the two-band model, the two sublattices in bulk originate from the change in the interchain bond length [53]. Thus, the epitaxial strain may restrict such a lateral structural change. As a result, in the film, the number of V-V dimerization merely reduces, whereas the fundamental electronic structures are maintained. In order to clarify this issue, a detailed structural analysis is necessary. The precise structural parameters for the V-V distance along the $c_R$ axis and V-O and V-V distances along the interchain would provide fruitful information on the puzzling phases in the $Cr_xV_{1-x}O_2$ films and pave the way for more realistic calculation incorporating the Peierls phenomena and electron-electron correlation.

## V. CONCLUSION

We determined the electronic phase diagram of $Cr_xV_{1-x}O_2$ films coherently grown onto $TiO_2$ (001) substrates. The phase diagram is similar to that of the bulk, whereas the $T_{MIT}$ is significantly suppressed by the epitaxial strain effect. Subsequently, we investigated the changes in the electronic structures and V-V dimerization of $Cr_xV_{1-x}O_2$ films via *in situ* PES and XAS measurements. The spectra exhibited remarkable changes as a function of $x$, which is in accordance with the transport properties, as follows. (1) The characteristic spectral changes associated with the cooperative Mott-Peierls MIT remain almost unchanged up to $x = 0.04$, whereas the hysteresis in the resistivity becomes small with increasing $x$. (2) The spectral change across the MIT remains intact even at $x = 0.08$, where there is no detectable hysteresis, whereas a pseudogap is formed at $E_F$ in the HT metallic phase. (3) The temperature-driven MIT is accompanied by V-V dimer formation for $0 \leq x \leq 0.08$, whereas the number of V-V dimerization reduces with increasing $x$. (4) The energy gap at the LT insulating phase is essentially unchanged with $0 \leq x \leq 0.08$, despite the suppression of V-V dimerization. The invariance of the energy gap with respect to $x$ suggests that the temperature-induced MIT in $Cr_xV_{1-x}O_2$ arises primarily from the strong electron correlations. In other words, the temperature-driven MIT in $Cr_xV_{1-x}O_2$ is the V-V dimerization-assisted Mott transition. Meanwhile, (5) the pseudogap at $x = 0.08$ eventually evolves to a full gap (Mott gap) at $x = 0.12$, which is consistent with the disappearance of the MIT in the electronic phase. This composition-derived MIT demonstrates that the Mott insulating phase without V-V dimerization is stabilized at $x > 0.08$ as a result of the superiority of the Mott instability over the Peierls one. The findings of this study provide significant insight into the complicated electronic phase diagram of $Cr_xV_{1-x}O_2$.

## ACKNOWLEDGMENTS

The authors are very grateful to H. Suzuki and T. Komeda for the useful discussions. The authors acknowledge S. Miyazaki, A. Wada, R. Hayasaka for their support in the experiments. This work was financially supported by a Grant-in-Aid for Scientific Research (No. 20KK0117, No. 21K20498, No. 22H01947, and No. 22H01948) from the Japan Society for the Promotion of Science (JSPS); CREST (JPMJCR18T1) from the Japan Science and Technology Agency (JST); and the MEXT Element Strategy Initiative to Form Core Research Center (JPMXP0112101001). T.K. acknowledges financial support from the Division for Interdisciplinary Advanced Research and Education at Tohoku University. N.H. acknowledges financial support from the Chemistry Personnel Cultivation Program of the Japan Chemical Industry Association. The work performed at KEK-PF was approved by the Program Advisory Committee (proposals 2018S2-004, 2022T001, and 2021S2-002) at the Institute of Materials Structure Science, KEK.

# Supplemental Material

## Electronic phase diagram of Cr-doped VO$_2$ epitaxial films studied by *in situ* photoemission spectroscopy

D. Shiga[1,2], X. Cheng[1], T. T. Kim[1], T. Kanda[1,2], N. Hasegawa[1], M. Kitamura[2], K. Yoshimatsu[1], and H. Kumigashira[1,2,*]

[1] *Institute of Multidisciplinary Research for Advanced Materials (IMRAM), Tohoku University, Sendai, 980–8577, Japan*

[2] *Photon Factory, Institute of Materials Structure Science, High Energy Accelerator Research Organization (KEK), Tsukuba, 305–0801, Japan*

[*]kumigashira@tohoku.ac.jp

# I. Sample characterization

## A. Surface morphology

The atomically flat surfaces of all the measured $Cr_xV_{1-x}O_2$/TiO$_2$ (001) film samples were confirmed by *ex situ* atomic force microscopy (AFM), as shown in Fig. S1. The AFM images show similar surface morphologies, irrespective of the film composition $x$. The root-mean-square (rms) values of the films' surface roughness $R_{rms}$ estimated from the AFM images were all less than 0.2 nm. The values are almost the same as that of the original TiO$_2$ substrate ($R_{rms}$ = 0.16 nm). The $R_{rms}$ values of the measured films were all less than the V-V dimer length (approximately 0.3 nm), indicating that these films were controlled to the scale of the V-V dimer length and that the smooth surface and interface were maintained as the film was grown to a thickness of ~8 nm.

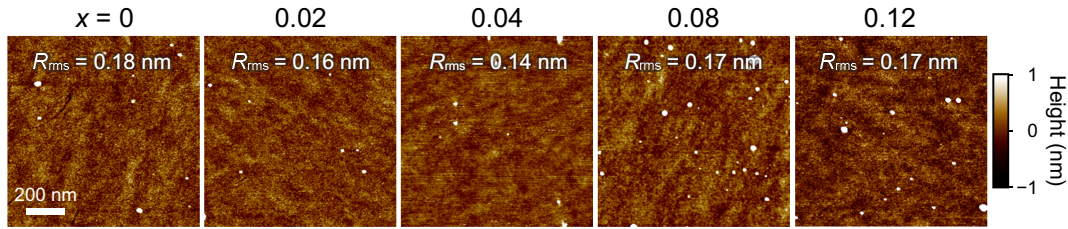

**Fig. S1.** Typical AFM images of the measured $Cr_xV_{1-x}O_2$/Nb:TiO$_2$ (001) films with $x$ = 0–0.12.

## B. Crystal structure

The crystal structures of the $Cr_xV_{1-x}O_2$ films were characterized by x-ray diffraction (XRD) measurements, which confirmed the achievement of a single phase in the $Cr_xV_{1-x}O_2$ films and the coherent growth of these films on TiO$_2$ (001) over the entire composition range of $0 \leq x \leq 0.12$. Figure S2(a) shows reciprocal space mapping (RSM) around the 112 reciprocal point. The RSMs confirm that a single-phase $Cr_xV_{1-x}O_2$ solid solution is formed over the entire composition range of $0 \leq x \leq 0.12$. The formation of single-phase $Cr_xV_{1-x}O_2$ film is also confirmed by the out-of-plane XRD patterns shown in Fig. S2(b), where the well-defined Laue fringes indicative of atomically flat surfaces and chemically abrupt interfaces are observed. As can be seen in Fig. S2(a), the in-plane ($a_R$ axis) lattice constant of $Cr_xV_{1-x}O_2$ films is pinned at that of TiO$_2$ substrates irrespective of $x$ [Fig. S2(b)], indicating the coherent growth of the $Cr_xV_{1-x}O_2$ films on $TiO_2$ substrates. Meanwhile, the out-of-plane ($c_R$ axis) lattice constant of the $Cr_xV_{1-x}O_2$ films increases with increasing $x$, reflecting a decrease in the mismatch between the films and substrates [40,41]. As summarized in Fig. S2(c), the out-of-plane $c_R$-axis lattice constant increases almost linearly, indicating that the lattice of the $Cr_xV_{1-x}O_2$ film obeys Vegard's law.

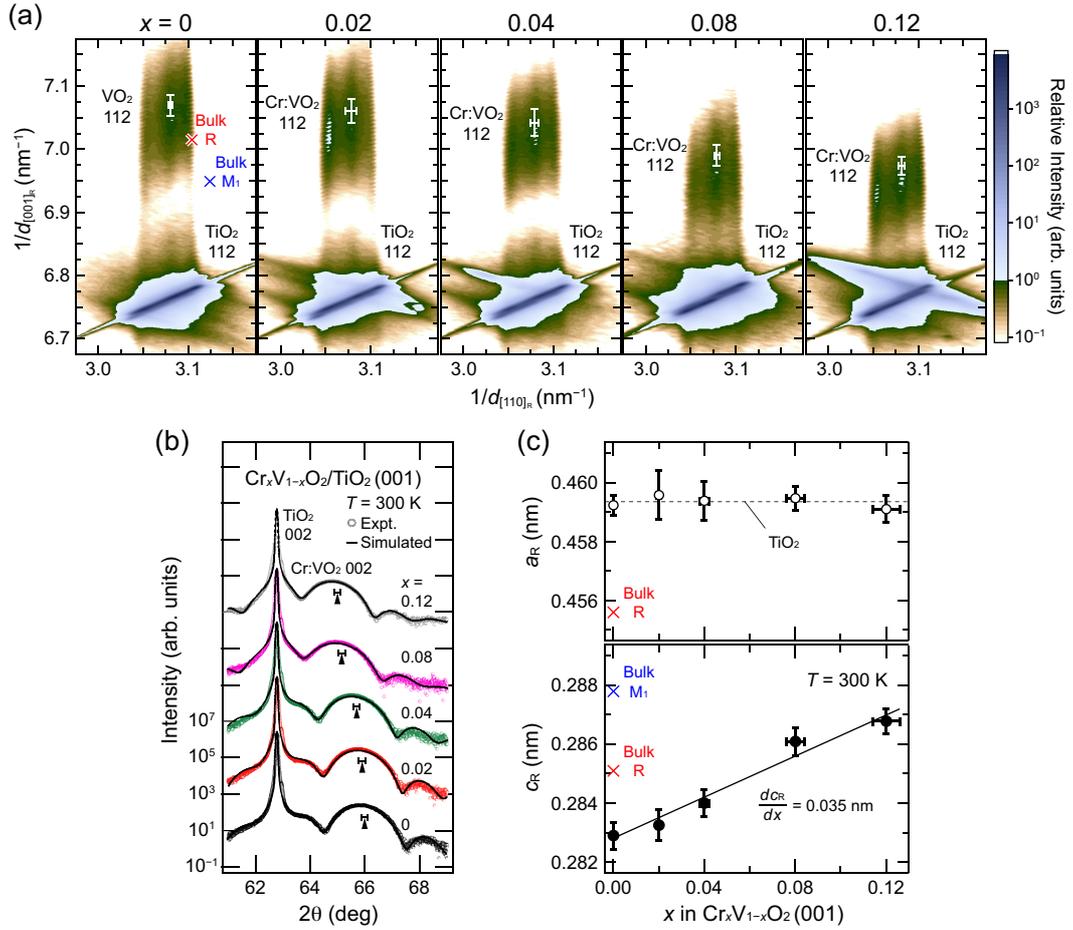

**Fig. S2.** (a) Typical RSMs around the 112 reciprocal point and (b) out-of-plane XRD patterns around 002 reflection for $Cr_xV_{1-x}O_2$ films on $TiO_2$ (001) substrates with $x$ = 0–0.12. (c) Plots of the $a_R$- and $c_R$-axis lattice constants as a function of $x$. For the $c_R$-axis lattice constants, the values estimated from the out-of-plane XRD patterns are plotted. The dashed line indicates the $a_R$ lattice constant of $TiO_2$ substrate. The $a_R$- and $c_R$-axis lattice constants for the rutile metallic (R) and monoclinic insulating ($M_1$) phases in bulk $VO_2$ ($x$ = 0) [40] are also shown in (a) and (c) for comparison.

## II. Experimental geometry in polarization-dependent x-ray absorption measurement

Figure S3 depicts the schematic of our experimental geometry for *in situ* polarization-dependent x-ray absorption spectroscopy (XAS) measurements, including the crystal axes of the $Cr_xV_{1-x}O_2$ (001) film sample and polarization vector ***E***. Regarding linear dichroism measurements for XAS, we acquired the spectra at angles of $\theta = 0°$ and $60°$ between the $a_R$-axis direction, which is defined as the *a*-axis direction in the rutile structure ($[100]_R$), and ***E*** while maintaining a fixed angle of $60°$ between the direction normal to the $(001)_R$ surface and the incident light. The maintenance of the fixed angle between the direction normal to the surface and the incident light ensures that the probing depth corresponding to the two spectra with different $\theta$ values is the same. In the present experimental geometry, XAS spectra with ***E*** // $c_R$ ($I_{//}$) can be deduced from the expression $I_{//} = (4/3)(I - I_\perp/4)$, where $I$ and $I_\perp$ (namely, that corresponding to ***E*** $\perp c_R$) denote XAS spectra measured with grazing ($\theta = 60°$) and normal ($\theta = 0°$) incidences, respectively.

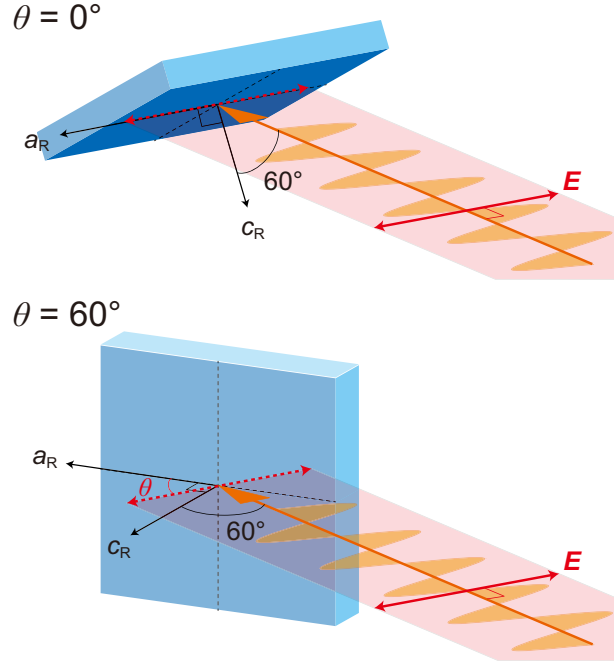

**Fig. S3.** Schematic of our experimental geometry of polarization-dependent XAS measurements of $Cr_xV_{1-x}O_2$ (001) films at $\theta = 0°$ and $60°$. The $a_R$- and $c_R$-axis directions are defined as the *a*- and *c*-axis directions in the rutile structure, respectively.

## III. Electronic structure of doped Cr ions

Figure S4(a) shows the Cr $L_{3,2}$-edge XAS (Cr $L$ XAS) spectrum measured at $T = 360$ K [high-temperature (HT) metallic phase] for the epitaxial $Cr_xV_{1-x}O_2$/Nb:TiO$_2$ (001) film at $x = 0.08$, in addition to those of Cr oxides [42] for comparison. The spectrum of the $Cr_xV_{1-x}O_2$ (001) film is almost identical to that of $Cr_2O_3$ ($Cr^{3+}$) and differs from that of $CrO_2$ ($Cr^{4+}$). Therefore, it is evident that Cr ions in $Cr_xV_{1-x}O_2$ films are trivalent.

To identify the contribution of Cr $3d$ states to the electronic structures near the Fermi level ($E_F$), we performed Cr $2p$–$3d$ resonant photoemission spectroscopy (PES). Figure S4(b) shows resonant PES spectra in the valence-band region measured at the HT metallic phase for the $Cr_xV_{1-x}O_2$ films at $x = 0.08$. The spectra were taken at the corresponding photon energies for the on ($h\nu = 577.7$ eV) and off ($h\nu = 573.9$ eV) resonances as indicated in Fig. 4(a). The Cr-$3d$ spectrum, which represents the Cr-$3d$ partial density of states, is obtained by subtracting the off-resonant spectrum from the on-resonant spectrum. The Cr-$3d$ spectrum exhibits a prominent single peak at a binding energy of 1.6 eV for the HT metallic phase. We observed almost the same spectral behavior for other $x$, as shown in Fig. S5. These results indicate that the doped Cr ions in $Cr_xV_{1-x}O_2$ films are trivalent, and their $3d$ states are localized at binding energies of 1.5–1.8 eV.

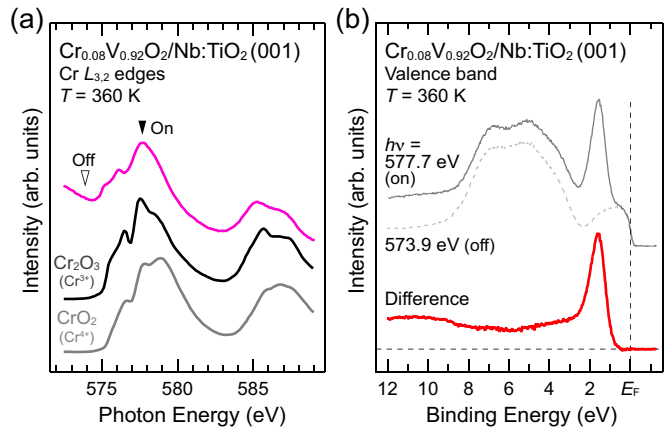

**Fig. S4.** (a) Cr $L$ XAS spectrum measured at $T = 360$ K for an epitaxial $Cr_{0.08}V_{0.92}O_2$/Nb:TiO$_2$ (001) film, in addition to those of $Cr_2O_3$ and $CrO_2$ [42] as references for the $Cr^{3+}$ and $Cr^{4+}$ states, respectively. The solid and open triangles denote the photon energies

used for the on- and off-resonant photoemission spectra shown in (b), respectively. (b) Cr 2p–3d resonant PES spectra taken at the corresponding photon energies for the on and off resonances. The Cr-3d spectrum (a thick red line) is obtained by subtracting the off-resonant spectrum (a gray dashed line) from the on-resonant spectrum (a gray solid line).

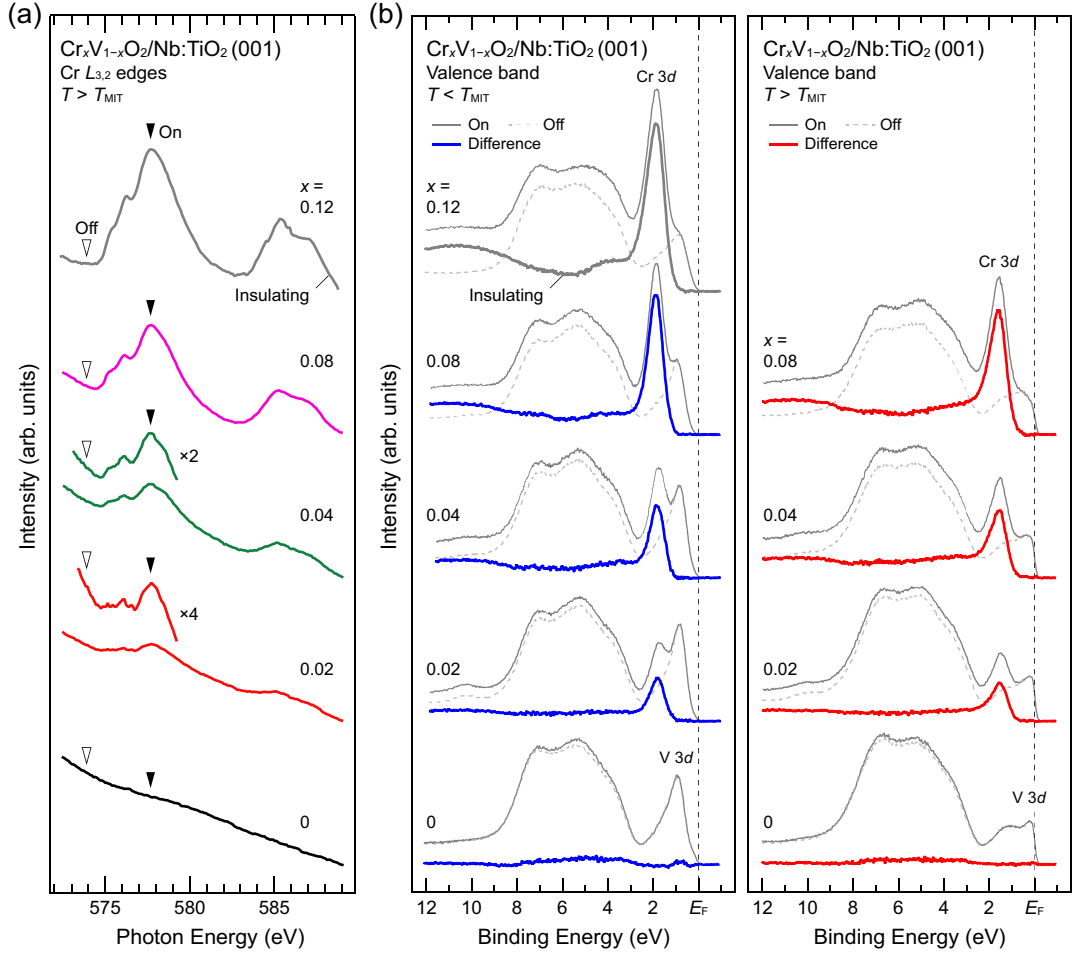

**Fig. S5.** (a) Cr $L$ XAS spectra measured in the HT metallic phase ($T > T_{MIT}$) for the $Cr_xV_{1-x}O_2$/Nb:TiO$_2$ (001) films with $x = 0$–0.08 and that in the insulating phase for $x = 0.12$. The solid and open triangles denote the photon energies used for the on- and off-resonant photoemission spectra shown in (b), respectively. (b) Cr 2p–3d resonant PES spectra taken at the corresponding photon energies for the on and off resonances for insulating (left panel) and metallic (right panel) phases. The Cr-3d spectra (thick-colored lines) are obtained by subtracting the off-resonant spectra (gray dashed lines) from the on-resonant spectra (gray solid lines).